\documentclass[11pt,dvips]{article}
\usepackage{epsfig,times} 

\usepackage{picinpar}
\usepackage{wrapfig}
\usepackage{floatflt}
\makeatletter
\renewcommand{\section}{\@startsection{section}{1}{0in}
	{0.4\baselineskip}{0.1\baselineskip}{\Large\bf}}
\renewcommand{\subsection}{\@startsection{subsection}{2}{0in}
	{0.25\baselineskip}{-\baselineskip}{\large\bf}}
\renewcommand{\subsubsection}{\@startsection{subsubsection}{3}{0in}
	{0.1\baselineskip}{-\baselineskip}{\normalsize\bf}}
\makeatother
\begin{document}

%
\thispagestyle{myheadings}
%
\markright{SH 3.2.42}
\begin{center}
%
{\LARGE \bf  Observation of the Moon Shadow in Deep Underground Muon Flux}
\end{center}

\begin{center}
%
%
{\bf J.H.~Cobb$^3$, M.L.~Marshak$^2$, W.W.M.~Allison,$^3$, G.J.~Alner$^4$, D.S.~Ayres$^1$,
W.L.~Barrett,$^6$, C.~Bode,$^2$, P.M.~Border$^2$, C.B.~Brooks$^3$, R.J.~Cotton$^4$,
H.~Courant$^2$, D.M.~Demuth$^2$, T.H.~Fields$^1$, H.R.~Gallagher$^3$, C.~Garcia-Garcia$^4$,
M.C.~Goodman$^1$, R.~Gran$^2$, T.~Joffe-Minor$^1$, T.~Kafka$^5$, S.M.S.~Kasahara$^2$,
W.~Leeson$^5$, P.J.~Litchfield$^4$, N.P.~Longley$^2$, W.A.~Mann$^5$, R.H.~Milburn$^5$,
W.H.~Miller$^2$, C.~Moon$^2$, L.~Mualem$^2$, A.~Napier$^5$, W.P.~Oliver$^5$, G.F.~Pearce$^4$,
E.A.~Peterson$^2$, D.A.~Petyt$^4$, L.E.~Price$^1$, K.~Ruddick$^2$, M.~Sanchez$^5$, J.~Sankey$^2$,
J.~Schneps$^5$, M.H.~Schub$^2$, R.~Seidlein$^1$, A.~Stassinakis$^3$, J.L.~Thron$^1$,
V.~Vassiliev$^2$, G.~Villaume$^2$, S.~Wakely$^2$, N.~West$^3$, D.~Wall$^5$\\
(The Soudan 2 Collaboration)}\\
{\it $^{1}$Argonne National Laboratory, Argonne, IL 60439, USA\\
$^{2}$University of Minnesota, Minneapolis, MN 55455, USA\\
$^3$Department of Physics, University of Oxford, Oxford OX1 3RH, UK\\
$^4$Rutherford Appleton Laboratory, Chilton, Didcot, Oxfordshire OX11 0QX, UK\\
$^5$Tufts University, Medford MA 02155, USA\\
$^6$Western Washington University, Bellingham, WA 98225, USA}
\end{center}

\begin{center}
{\large \bf Abstract\\}
\end{center}
\vspace{-0.5ex}

A shadow of the moon, with a statistical significance of $5\sigma$, has been observed
in the underground muon flux at a depth of 2090 mwe using the Soudan 2 detector. The
angular resolution of the detector is well described by a Gaussian with $\sigma \le
0.3^{\circ}$. The position of the shadow confirms the alignment of the detector to
better than $0.15^{\circ}$. This alignment has remained stable during 10 years of data taking
from 1989 through 1998.
%

\vspace{1ex}

\section{Introduction}
\label{intro.sec}
The Soudan 2 detector, a 963 tonne iron and drift-tube sampling calorimeter designed to
search for nucleon decay (Allison, 1996), is located in the Soudan Mine in 
northeastern Minnesota, USA, at $47.8^{\circ}$ W, $92.3^{\circ}$ N. The detector has 
recorded $>5 \times 10^7$ deep underground muon tracks
during the entire ten-year interval from January 1989 to December 1998. These events
provide a rich data source in which to search for cosmic ray muon point sources, provided
that the angular resolution and pointing accuracy of the detector can be well
understood.

As suggested by Clark (1957), the cosmic ray shadows of the moon and sun test
both the angular resolution and pointing of a cosmic ray detector. The angular diameter
of both bodies as observed at the earth is $0.5^{\circ}$. Observation
of a shadow places limits on detector angular resolution and pointing
accuracy, as well as on phenomena affecting cosmic ray propagation and interaction. 
These latter effects include deflections due to the geomagnetic field and,
in the case of the sun, the solar and interplanetary magnetic fields, and smearing due
to multiple Coulomb scattering and production mechanisms for air showers and muons.
Overall, detection of a cosmic ray shadow is easier for higher rather than lower
energy cosmic rays and easier for the moon than for the sun, 
because of the time variability of the solar and
interplanetary magnetic fields. Several large air shower arrays have 
previously reported observation of lunar and solar shadows at primary energies
ranging from 10 to 100 TeV (Alexandreas, 1991; Borione, 1994; Merck, 1996;
Amenomori, 1996). The MACRO detector has also reported the observation
of the lunar cosmic ray shadow with deep underground muons (Ambrosio, 1999). The angular
resolution for the MACRO detector derived from this observation is 
$\sigma \approx 0.9^{\circ}$.

We report here on the observation of the lunar shadow in the Soudan 2 deep
underground muon data during the interval 1989 to 1998 and elsewhere on the
solar shadow during the same interval (Allison, 1999). The expected shadow deviation
and broadening can be calculated using Monte Carlo techniques. We have performed
a simulation study of 14,000 muon events from cosmic ray primaries passing within
$3^{\circ}$ of the moon using \textsc{HEMAS}, \textsc{SIBYLL} and \textsc{GEANT} cascade,
hadronic interaction and transport codes. The geomagnetic field was modelled
as a pure dipole, with a field vector at Soudan set at the observed 
magnitude of $5.9 \times 10^{-5}$ T. The minimum muon surface energy to penetrate to the
Soudan 2 detector depth from the direction of the moon is $0.8$ TeV. The simulation shows that
the mean energy of primaries coming from the direction of the moon which create muons at the
Soudan 2 detector depth is 19 TeV. The mean tranverse momentum ($\Delta \vec{P_t}$)
due to the geomagnetic field in the impulse approximation is 25 GeV/c. 
The expected mean geomagnetic deflection of the shadow center  
is $0.076^{\circ}$ to the west. The expected smearing of the lunar shadow is
highly non-Gaussian with tails in the deflection distribution extending beyond
$2^{\circ}$. More relevant for comparison with observed data 
are the 50th percentile angles of $0.2^{\circ}$ for geomagnetic deflection
and $0.3^{\circ}$ for multiple Coulomb scattering.

\section{Data Collection and Analysis}
\label{data.sec}

Events in the Soudan 2 detector are recorded upon satisfaction of a local
energy deposition trigger requirement. At that time, pulse heights from all detector
channels above threshold are digitized and recorded at $160 ns$
intervals for a time long enough to include the maximum possible electron drift of
50 cm. During offline analysis, muon track events are differentiated from other events
that result primarily from electronic noise or radioactivity. Muon tracks are reconstructed
using two different software algorithms. If both algorithms provide satisfactory
reconstructions, the directionality from the algorithm that provides the best angular
resolution is used. The reconstructed events are subjected to
both aggregate run cuts on parameters such as good event fraction and drift velocity
and individual events cuts on parameters such as track length and measured pulse height.
The goals of these cuts are to maximize angular resolution and minimize non-muon-track
contamination while not excessively reducing the numbers of events. For the moon shadow
data sample, a 100 cm minimum track length was required to select an event sample with
good angular resolution. For multimuon events, the parameters 
of the longest track were used to determine the directionality of the event. 
The total reconstructed event data sample passing cuts consisted of $3.4 \times 10^7$ muon events.

The time of each event is recorded using a time base synchronized to the {\sc WWVB}
time standard. The event time and the known detector coordinates specify the apparent
direction of the moon, including the correction for parallax. 
The angle $\theta$ between each muon track and
the direction of the moon at the time of the event is then calculated.
Fig. 1(a) shows a plot of the angular density of muons,
$(1/\pi)(dN_{\mu}/d\theta^2)$ vs. $\theta$, the angular distance between the
muon direction and the calculated position of the center of the moon. 
In the absence of a moon shadow,
this plot should be flat, because the varying direction of the moon averages over any
anisotropies in detector acceptance or rock overburden. The plot, however, clearly
shows a deficit of events at small angles, which we attribute to a lunar cosmic ray shadow.
The significance of the shadow is tested by comparing the difference in $\chi^2$
between the best fit to a flat distribution ($\chi^2 = 82.9$) and the best fit 
($\chi^2 = 58.3$) to the form
\begin{equation}
\frac{dN_{\mu}}{d\theta^2} =  \lambda (1 - \pi(R_m^2/2\sigma^2)exp(
-\theta^2/2\sigma^2))
\end{equation}
where $R_m = 0.26^{\circ}$ is the mean angular radius of the moon. $\lambda$ represents
the angular density of muons and $\sigma$ folds together the the angular resolution
and directional alignment of the detector, geomagnetic deflections, 
shower and muon production effects and the finite angular size of the moon. 
The best fit parameters are $\lambda
= 607 \pm 3.5$ muons per square degree and $\sigma = 0.33^{\circ} \pm 0.05^{\circ}$.
The value of $\lambda$ implies that the moon obscures a total of 129 muons,
about one per month. The improvement in the $\chi^2$ of 24.6 for 2 df implies
a chance probability of the observed shadow is $< 10^{-5}$, a statistical
significance of $5\sigma$.

To further test the angular resolution and alignment of the Soudan 2 detector, we have
constructed a two-dimensional contour map of the muon flux from the direction of the moon.
Cobb (1999) provides details of the procedure used for making this map. 
A summary of the process is as follows:

\begin{figwindow}[1,r,%
{\mbox{\epsfig{file=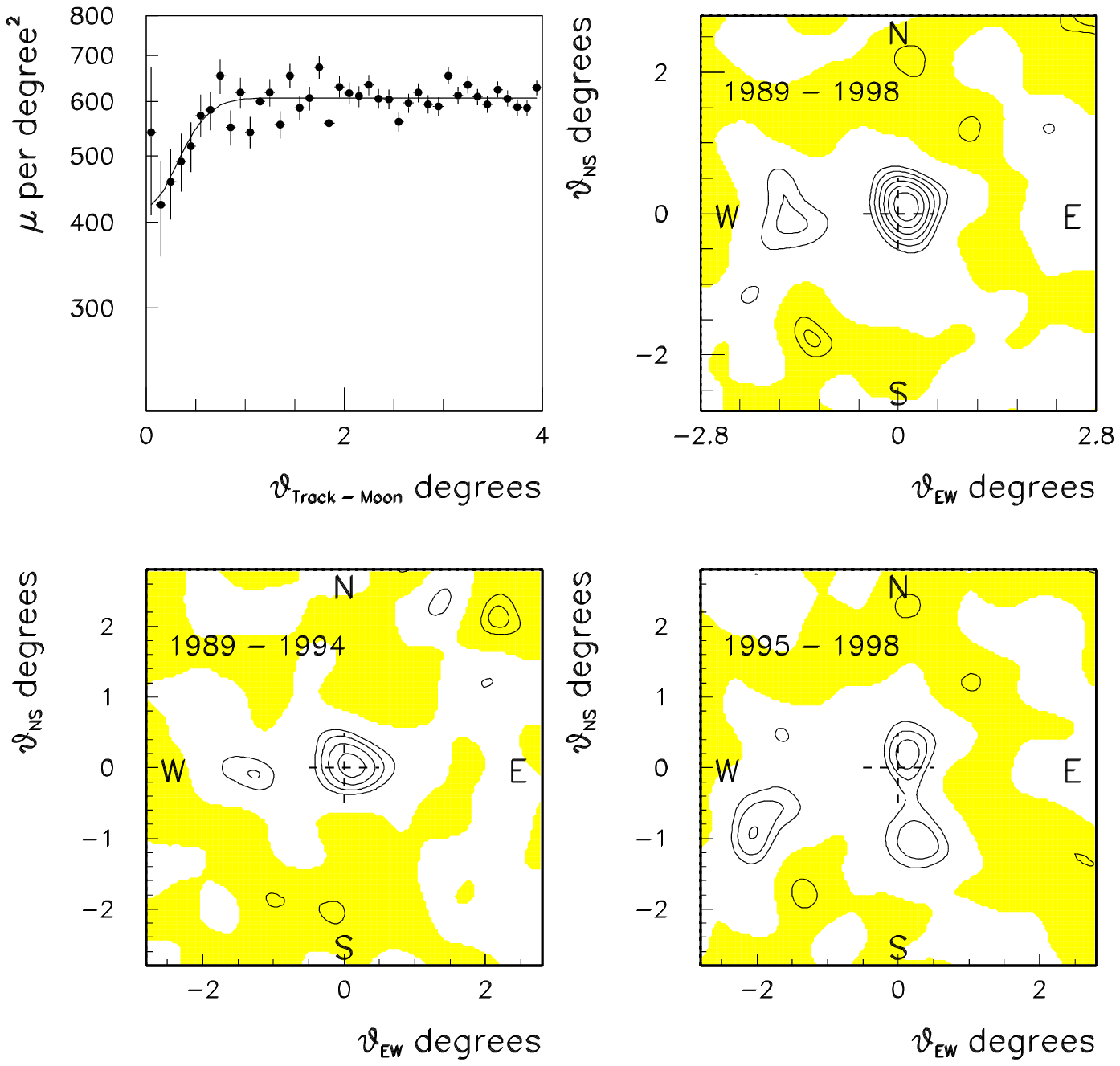,width=6.6in}}},%
{(a) (upper left) The angular density of muons, $(1/\pi)(dN_{\mu}/d\theta^2)$ 
vs. $\theta$, the angular distance between the
muon direction and the calculated position of the center of the moon. (b)
(upper right) Contour map of the normalized deviations, $Z$, for a $\pm2.8^{\circ}
\times \pm2.8^{\circ}$ region centered on the moon with a rebinning kernel
$\sigma_k = 0.29^{\circ}$. The contour lines are spaced by $\Delta Z = 0.5$
and are shown only where $|Z| \ge 2.0$. Regions with $Z > 0$ are shaded.
(c) (lower left) Same as (b) for the years 1989 to 1994. (d) (lower right)
Same as (b) for the years 1995 to 1998. }]
\end{figwindow}

The difference vector was calculated between the vector representing the muon
trajectory and the vector pointing to the detector from the moon at the muon's
arrival time. The projections of this difference vector in the east-west
and north-south planes, $\theta_{EW}$ and $\theta_{NS}$, were then calculated.
Muon events were binned in a two-dimensional array in $0.04^{\circ}$ by $0.04^{\circ}$
bins. A second similar two-dimensional histogram for the background or
expected number of events in each bin was then calculated by assuming factorization.
This hypothesis assumes that the number of events in each bin $dN_{\mu} =
\lambda d\theta_{EW}\theta_{NS} = Y(\theta_{EW}) \times Z(\theta_{NS}) d\theta_{EW}\theta_{NS}$.
The functions $Y(\theta_{EW})$ and $Z(\theta_{NS})$ were determined by fitting
quadratics to the $\theta_{EW}$ and $\theta_{NS}$ projections of the observed
event histogram. The next step was to smooth both the observed and expected event
histograms with a two-dimensional Gaussian kernel $w_{l,m} = (1/2\sigma_{k}^2)
exp(-\psi_{l,m}^2/2\sigma_{k}^2) \Delta\theta_{EW}\Delta\theta_{NS}$, where 
$\psi_{l,m}$ is the angular distance between bin $i,j$ and bin $l,m$ and 
$\sigma_k$ is a smoothing parameter. Finally, maps were made of the normalized
statistic $Z_{i,j} = (d_{i,j} - b_{i,j})\sqrt{var(b_{i,j})}$, where $d_{i,j}$
are the bin contents in the smoothed observed event histogram and $b_{i,j}$
are the bin contents in the smoothed expected event histogram. The density
of events per bin reported here is sufficient that $Z_{i,j}$ is approximately
normally distributed and confidence levels can be extracted by the usual tests
for Gaussian statistics. (See Cobb (1999) for references.)

Fig. 1(b) shows the resulting contour map for the entire ten-year data sample
with $\sigma_k$ chosen as $0.29^{\circ}$ in order to minimize $Z_{i,j}$ at the
center of the shadow. That minimum value of $Z_{i,j} = -4.98$
is located at $\theta_{EW} = 0.1^{\circ}$,
$\theta_{NS} = 0.1^{\circ}$, $i.e.$, within $0.15^{\circ}$ of the calculated
direction of the moon. The $Z = -4.5$ contour, which corresponds to $\approx
75\%$ CL, includes the origin. Thus, Fig. 1(b) suggests neither evidence for any
misalignment of the Soudan 2 detector nor any indication of a shadow offset
due to the geomagnetic field at the level of $\approx 0.1^{\circ}$. This latter observation
agrees with results of the Monte Carlo simulations described earlier.
The other contours shown in the figure are spaced at intervals in $Z_{i,j}$ of
0.5 and are only shown for $|Z_{i,j}| \ge 2.0$. The shaded areas indicate bins
with an excess of observed events over expected events. Except for the deep shadow
at the position of the moon, Fig. 1(b) appears as expected as a result of
statistical fluctuations.

The Soudan 2 detector consists of 216 modules and has changed considerably
in size and in the location of individual modules during the ten years of data
collection. The moon shadow tests whether the alignment
and calibration has remained constant during the entire data collection interval.
Figs. 1(c) and 1(d) show similar maps to the one in Fig. 1(b), except that Fig. 1(c)
represents data collected during 1989 through 1994 and Fig. 1(d) represents data
collected during 1995 through 1998. The moon shadow is seen clearly in both maps.
The position of the minimum is at $\theta_{EW} = 0.1^{\circ}$,
$\theta_{NS} = 0.02^{\circ} ~(Z = -3.78)$ for 1989-1994 and $\theta_{EW} = 0.1^{\circ}$,
$\theta_{NS} = 0.18^{\circ} ~(Z = -3.39)$ for 1995-1998. The depths of both minima
are close to the expected value of $Z = -4.98/\sqrt{2} = -3.52$ expected for
half the total sample. The positions of these shadows confirm that the detector
alignment has remained stable over the entire data collection decade.

\section{Conclusions}
\label{conclusions.sec}

A $5\sigma$ shadow of the moon in the underground muon flux has been observed using the
Soudan 2 detector. The shadow position confirms that the alignment of the detector 
is correct to  $\le 0.15^{\circ}$ and that the angular resolution of the detector (including
geomagnetic dispersion, shower and muon production, multiple Coulomb scattering,
and the finite size of the moon)
can be adequately described by a Gaussian point spread function with $\sigma
= 0.29^{\circ}$. The division of the data sample into two parts which display
nearly identical shadows suggests that the alignment and resolution 
of the detector has been stable over the entire ten-year data collection interval.  

%
\vspace{1ex}
\begin{center}
{\Large\bf References}
\end{center}
%
Alexandreas,~D.E.,~{\it et al.}, 1991 Phys. Rev. \textbf {D43}, 1735.\\
Allison,~W.W.M.,~{\it et al.}, 1996 Nucl. Inst. and Meth. \textbf {A376},
 377 and \textbf {A381}, 385.\\
Allison,~W.W.M.,~{\it et al.}, 1999 ICRC Paper SH 3.2.43.\\ 
Ambrosio,~M., ~{\it et al.}, 1999 Phys. Rev. \textbf {D59}, 012003.\\
Amenomori,~M.,~{\it et al.}, 1996 Astrophys. J. \textbf {464}, 954.\\
Borione,~A., ~{\it et al.}, 1994 Phys. Rev. \textbf {D49}, 1171.\\
Clark,~G.W., 1957 Phys. Rev. \textbf {108}, 450.\\
Cobb,~J.H., ~{\it et al.}, 1999, submitted to Phys. Rev. D.\\
Merck,~M., ~{\it et al.}, 1996 Astropart. Phys. \textbf {5}, 379.\\

\end{document}